\documentclass[twocolumn,pre,floatfix,superscriptaddress]{revtex4}
\usepackage[T1]{fontenc}
\usepackage[latin9]{inputenc}
\usepackage{mathrsfs}
\usepackage{amsmath}
\usepackage{amssymb}
\usepackage{graphicx}
\usepackage{braket}
\usepackage{color}
\usepackage[svgnames]{xcolor}
\usepackage{comment}
\usepackage{natbib}
\usepackage{dsfont}
\usepackage{mdframed}

\makeatletter

\usepackage{listings}

\usepackage{hyperref}

\hypersetup{
  colorlinks=true,
  linkcolor=blue,
  citecolor=ForestGreen,
  urlcolor=DarkOrchid
}

\AtBeginDocument{
  
}

\makeatother

\usepackage{babel}

\begin{document}

%\title{Combinatorial Optimization with Physics-Inspired Graph Neural Networks - Reply to Angelini and Ricci-Tersenghi}
\title{Reply to: Modern graph neural networks do worse than classical greedy algorithms in solving combinatorial optimization problems like maximum independent set}

\author{Martin~J.~A.~Schuetz}
\affiliation{Amazon Quantum Solutions Lab, Seattle, Washington 98170, USA}
\affiliation{AWS Center for Quantum Computing, Pasadena, CA 91125, USA}

\author{J.~Kyle~Brubaker}
\affiliation{Amazon Quantum Solutions Lab, Seattle, Washington 98170, USA}

\author{Helmut G.~Katzgraber}
\affiliation{Amazon Quantum Solutions Lab, Seattle, Washington 98170, USA}
\affiliation{AWS Center for Quantum Computing, Pasadena, CA 91125, USA}

\date{\today}

\begin{abstract}

% Reply to Maria Chiara Angelini \& Federico Ricci-Tersenghi Nature Machine Intelligence [DOI xxx] (2022). 
We provide a comprehensive reply to the comment written by Chiara Angelini and Federico Ricci-Tersenghi [arXiv:2206.13211] and argue that the comment singles out one particular non-representative example problem, entirely focusing on the maximum independent set (MIS) on \textit{sparse} graphs, for which greedy algorithms are expected to perform well. Conversely, we highlight the broader algorithmic development underlying our original work \citep{schuetz:22}, and (within our original framework) provide additional numerical results showing sizable improvements over our original results, thereby refuting the comment's performance statements. We also provide results showing run-time scaling superior to the results provided by Angelini and Ricci-Tersenghi. Furthermore, we show that the proposed set of random $d$-regular graphs does not provide a universal set of benchmark instances, nor do greedy heuristics provide a universal algorithmic baseline. Finally, we argue that the internal (parallel) anatomy of graph neural networks is very different from the (sequential) nature of greedy algorithms and emphasize that graph neural networks have demonstrated their potential for superior scalability compared to existing heuristics such as parallel tempering. We conclude by discussing the conceptual novelty of our work and outline some potential extensions. 

\end{abstract}

\date{\today}

\maketitle

\textbf{Problem instances and benchmarks.} 
The comment by Angelini and Ricci-Tersenghi is exclusively focused on the maximum independent set (MIS) problem for \textbf{sparse} random $d$-regular graphs with low densities between $\sim 10^{-6}$ and $\sim 10^{-3}$, in line with one of the comment's main references entitled ``\textit{Monte Carlo algorithms are very effective in finding the largest independent set in \textbf{sparse} random graphs}.'' However, the comment leaves out the fact that we have also provided results for standard MaxCut benchmark instances based on the publicly-available (and dense) \texttt{Gset} data set, as provided in Table I of our paper \citep{schuetz:22}. We report on a wide array of benchmark results, including results based on (i) an SDP solver using dual scaling (DSDP), (ii) Breakout Local Search (BLS), (iii) a Tabu Search metaheuristic (KHLWG), and (iv) a recurrent graph neural network (GNN) architecture for maximum constraint satisfaction problems (RUN-CSP). We find that our simple graph convolutional network (GCN) baseline architecture is competitive with these solvers and typically within approximately 1\% of the best results based on BLS. In addition, we have chosen to provide results for random $d$-regular results primarily because these allow one to perform large-scale experiments and averaging over instances, as well as comparisons to analytical bounds \citep{duckworth:09}. At large scales these instances are sparse, and it is not surprising that greedy algorithms perform well in this regime. As such, applying our approach to merely sparse random $d$-regular graphs is, indeed, overkill. However, our GNN-based approach is more broadly applicable. Here we explicitly disagree with the comment, because our approach is not inherently limited to \textit{sparse} graphs only.  Specifically, we refer to our follow-up work on graph coloring problems where we have presented results demonstrating that physics-inspired graph neural networks (PI-GNNs) show the potential to outperform greedy algorithms \citep{schuetz:22b}, in particular for \textbf{dense} instances.  Moreover, it will arguably always be possible to design a better heuristic if one restricts the analysis to a specific problem (such as MIS) on special instances (such as sparse $d$-regular graphs). However, going beyond sparse $d$-regular graphs, we still think that the Boppana-Halldorsson (BP) algorithm, just as the Goemans-Williamson (GW) algorithm for MaxCut, were reasonable algorithmic choices to compare to. These are fairly established, widely-used algorithms and we have shown on-par performance, with much better scalability. We do not think that greedy algorithms---while performant for sparse instances---provide a universal baseline, as evident from our results for dense instances presented in Ref.~\citep{schuetz:22b}. Other heuristics mentioned by the authors, such as simulated annealing or parallel tempering Monte Carlo, can typically be not scaled to system sizes with millions of variables, let alone billion-scale problems that GNNs can tackle \citep{zheng:20}. We do not claim to outperform {\em all} heuristics. In fact, that is arguably not possible for any (meta) heuristic that, by definition, cannot provide a universal, provable speed-up. 

\textbf{Proposed benchmark instances.} 
Following their focus on random $d$-regular graphs, the authors make a case for benchmark instances based on random $d$-regular graphs with $d>16$. While this is certainly an interesting academic exercise, we are not convinced about its practical usefulness, simply because typically real-world problems are highly structured, as opposed to the proposed random instances. For example, for $d$-regular graphs the degree distribution is fixed at $d$ (every node is connected to $d$ other nodes) whereas real networks are known to be heavily skewed with the degree distribution showing a long tail of values that are far above the mean \citep{leskovec}. Similarly, many real-world networks are known to contain hubs of highly connected nodes---such as social networks \citep{leskovec, monti:19}. Such features are not captured by random $d$-regular graphs, thus potentially limiting their usefulness as a guideline for real-world applications. 

\begin{figure}
\includegraphics[width=1.0 \columnwidth]{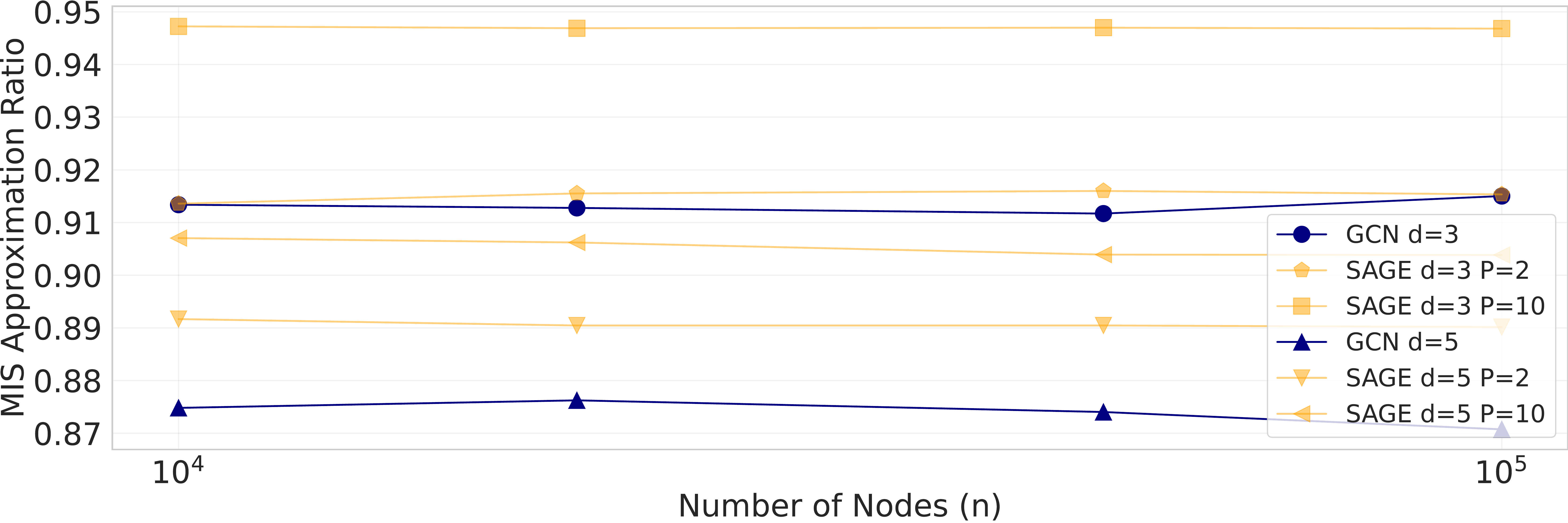}
\caption{
Size of the independent set for random $d$-regular graphs, as a function of the number of nodes $n$, reported as relative approximation ratio (AR) with respect to the known theoretical upper bounds. In addition to the GCN-based results (dark, as previously reported) numerical (average) results are shown for a GraphSAGE architecture (light), for both $P=2$ (as used previously) and $P=10$. GraphSAGE consistently provides slightly larger independent sets. Further improvements can be achieved by simply setting $P=10$.  All data points are averaged over $20$ samples per problem size. 
\label{fig:mis_sage}}
\end{figure}

\textbf{Algorithmic development}. 
The comment by Angelini and Ricci-Tersenghi equates the proposed GNN architecture with the simple, vanilla GCN used in this work for demonstration purposes. However, graph neural networks represent an entire family of neural networks under the larger umbrella of geometric deep learning \citep{bronstein:21}. Therefore, there is not just one GNN architecture just as there is not just one genetic algorithm, but rather a whole family of GNN-based architectures within one larger framework. In our original numerical experiments we have restricted ourselves to a simple two-layer GCN architecture as it represents a simple baseline model. However, the conceptual novelty is much broader. To demonstrate this, we have performed additional numerical experiments for the MIS problem on $d$-regular graphs with $n \in [10^4,10^5]$. The results are shown in Fig.~\ref{fig:mis_sage}. First, by simply replacing the GCN module with GraphSAGE (which amounts to changing just a few lines of code) we observe a consistent improvement in the average independence number. This improvement is more pronounced for denser $d=5$ graphs, in line with our results reported in Ref.~\citep{schuetz:22b}. Second, in all experiments we have set the penalty parameter $P=2$ for simplicity. However, the penalty term can be further refined in a simple outer loop. For example, by simply setting $P=10$ we find even further consistent improvements. Specifically, for $d=3$ at $P=10$ GraphSAGE achieves approximation ratios of $\mathrm{AR} \sim 0.947$, on par with the DGA-based results reported by Angelini and Ricci-Tersenghi. One may still see a small performance gap for $d=5$, but that is arguably because we used the same hyperparameters as used previously for $P=2$. Further approaches for potential improvements are discussed below. Given the improvements seen already with these two simple tweaks, we emphasize that the claims made in the comment do not hold in generality. 

Overall, the authors show $\sim 5\%$ performance improvements compared to our simple PI-GCN baseline results. This comparison is against a given vanilla GNN architecture which can be easily improved upon, as shown above. Thus, we think that the comment---entirely focusing on MIS and on sparse instances---as well as a vanilla GCN architecture singles out a particular non-representative example. Finally, we point out that conceptually the comparison to greedy algorithms is somewhat misleading; in fact, the internal anatomy of PI-GNN is very different from greedy algorithms, because updates to node representations are done in a fully parallelized fashion, as opposed to the sequential nature of greedy methods. 

\begin{figure}
\includegraphics[width=1.0 \columnwidth]{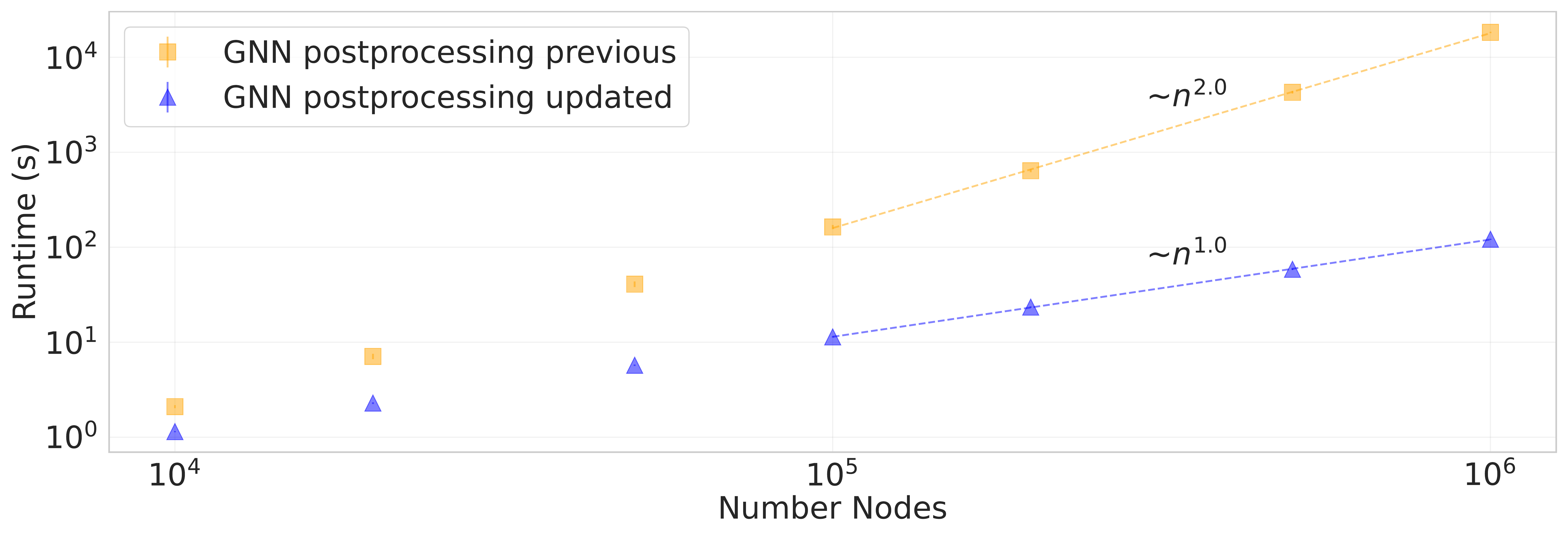}
\includegraphics[width=1.0 \columnwidth]{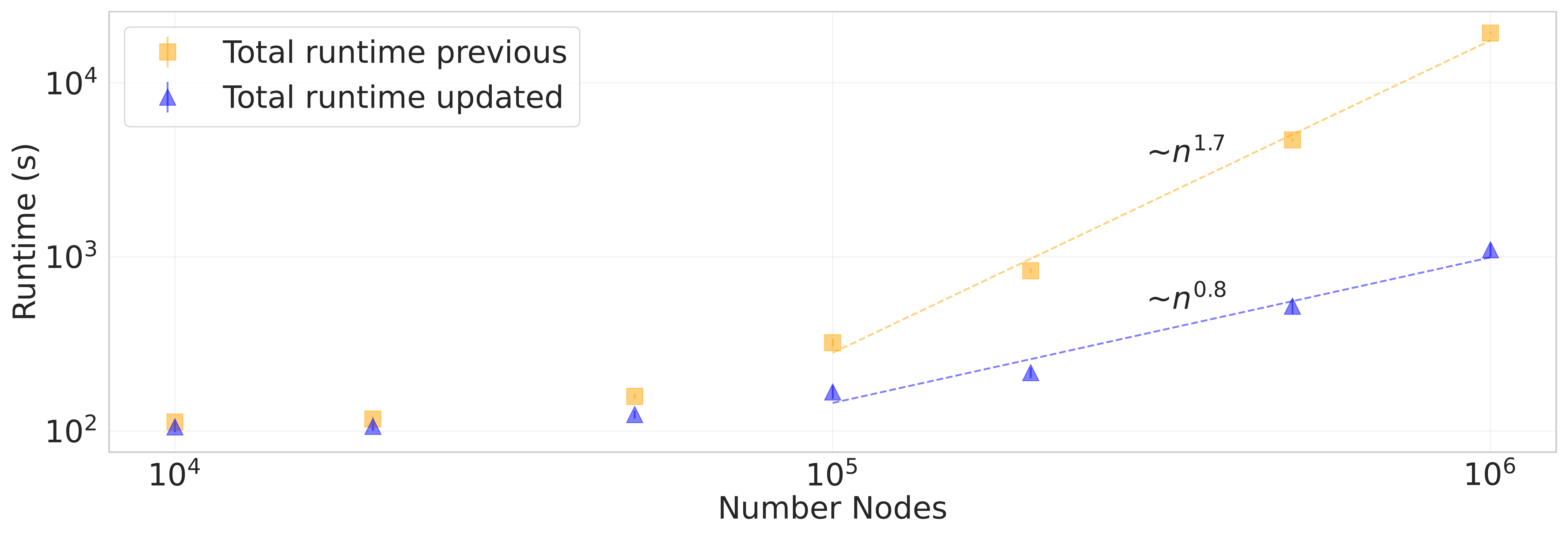}
\caption{
Run time in seconds for MIS problems on random $3$-regular graphs as a function of the number of nodes $n$. \textbf{Upper panel}: Post-processing times required to check for constraint violations of the independence condition. In addition to our previous post-processing routine (orange squares) we provide results for a simplified routine (blue triangles) that shows a simple linear scaling with $n$ (and thus linear with the number of edges for $d$-regular graphs). \textbf{Lower panel}: Total PI-GNN run time involving both the GNN training time and post-processing. With the updated post-processing routine, the aggregate run time is dominated by the GNN model training, displaying sub-linear run time approximately scaling as $\sim n^{0.8}$, as previously reported. All data points are averaged over $20$ samples per problem size. 
\label{fig:runtimes}}
\end{figure}

\textbf{Runtimes.} 
The comment also criticizes the GNN's run time as quadratic in the number of nodes $n$. As noted already in our paper \citep{schuetz:22}, the GNN model training alone displays sub-linear run time scaling as $\sim n^{0.8}$, in line with our MaxCut results, while only the {\em aggregate} run time (including post-processing to enforce the independence condition) is slower to make sure no constraints are violated. Unfortunately, the comment does not mention this fact, as well as the fact that we have previously stated that ``\textit{\ldots we have observed these violations only in very few cases\ldots}.''  Just like for MaxCut, the actual solver run time is approximately linear (and not quadratic). In fact, for $P=10$ we have not observed any constraint violations, i.e., there is no overhead. Still, we have revisited and simplified our post-processing routine. While our previous logic checked for overlap between the graph's edge set and the edges associated with the complete subgraph involving all marked vertices (an effort quadratic in the number of marked vertices), our revised routine simply scans through all edges (a linear effort for $d$-regular graphs) and checks if the nodes connected by this edge are both marked. The results are shown in Fig.~\ref{fig:runtimes}. We find that, with this updated post-processing routine, the aggregate run time is dominated by the  GNN model training, displaying sub-linear run time scaling as $\sim n^{0.8}$, as previously reported. Thus, we conclude that the scaling of the run time with the problem size $n$ observed with PI-GNN is superior to the DGA-based scaling $\sim n^{1.15}$ reported by Angelini and Ricci-Tersenghi. Furthermore, we point out that in the future techniques such as transfer learning  (as commonly applied across deep learning) could further boost the time to solution achieved with PI-GNN.

\textbf{Prospects for solving combinatorial optimization problems with neural networks.} 
The authors are pessimistic that neural networks could provide any advantages over existing methods, for both sparse as well dense instances. We think that this view is overly pessimistic for the following reasons. First, again we refer to Ref.~\citep{schuetz:22b} where we have already presented results demonstrating that GNNs show the potential to outperform existing algorithms, in particular for larger, denser instances. In addition, we refer to Ref.~\citep{fan:21} where extensive calculations on two-, three-, and four-dimensional Edwards-Anderson spin glass instances have been performed, showing superior performance of a neural network based approach over existing methods (in fact, including both a greedy algorithm as well as parallel tempering). Second, the graph neural network referred to by the authors in their Ref.~[15] is based on a vanilla GCN architecture which can likely be improved upon, as shown above. Third, GNNs have been used successfully on billion-scale graphs \citep{zheng:20}, while existing heuristics such as parallel tempering are typically limited to problems with $\sim 10^4$ variables at most, thus demonstrating their potential for superior scalability. Finally, with reference to message passing algorithms, we point out that the defining feature of many existing GNN architectures is that they use some form of neural message passing in which vector messages are exchanged between nodes and updated using neural networks \citep{gilmer:17}. 

\textbf{Conceptual novelty.} 
Overall, we think that the comment by Angelini and Ricci-Tersenghi is limited in scope, with claims that do not hold in general, and does not pay appropriate attention to the conceptual novelty of our work. The comment is exclusively focused on the MIS problem on sparse instances, whereas we outline broader applicability to other problems, within the large family of quadratic unconstrained and polynomial unconstrained binary problems. In addition, we recently showed that the PI-GNN framework can be readily extended to entirely new problem classes, such as the graph coloring problem \citep{schuetz:22b}. As mentioned already in the outlook section of our paper, we believe that our work can motivate and trigger a vast array of interesting follow-up studies, including but not limited to more detailed analyses of the limitations of the proposed GNN-based framework. For example, we have already shown how simple tweaks of the GNN architecture can result in sizable performance improvements \citep{schuetz:22b}. Apart from GraphSAGE, alternative candidates are Graph Attention Networks (GATs) or Graph Isomorphism Networks (GINs) \citep{xu:19}, among others. In addition, beyond the baseline approach we chose in our original paper \citep{schuetz:22}, more sophisticated training schemes (for example, incorporating ideas from annealing), graph rewiring strategies to decouple the GNN training graph from the problem graph \citep{topping:21}, with the potential to break locality, schemes to incorporate structural graph features into the node embeddings, as well as more refined post-processing techniques should all help improve the performance of the proposed GNN approach. 

{\bf Competing interests.} M.J.A.S., J.K.B. and H.G.K. are listed as inventors on a US provisional patent application (no. 7924-38500) on combinatorial optimization with graph neural networks.

\textbf{Correspondence and requests for materials} should be addressed to Martin J. A. Schuetz, J. Kyle Brubaker or Helmut G. Katzgraber.

\bibliography{refs}

\begin{thebibliography}{11}
\expandafter\ifx\csname natexlab\endcsname\relax\def\natexlab#1{#1}\fi
\expandafter\ifx\csname bibnamefont\endcsname\relax
  \def\bibnamefont#1{#1}\fi
\expandafter\ifx\csname bibfnamefont\endcsname\relax
  \def\bibfnamefont#1{#1}\fi
\expandafter\ifx\csname citenamefont\endcsname\relax
  \def\citenamefont#1{#1}\fi
\expandafter\ifx\csname url\endcsname\relax
  \def\url#1{\texttt{#1}}\fi
\expandafter\ifx\csname urlprefix\endcsname\relax\def\urlprefix{URL }\fi
\providecommand{\bibinfo}[2]{#2}
\providecommand{\eprint}[2][]{\url{#2}}

\bibitem[{\citenamefont{Schuetz
  et~al.}(2022{\natexlab{a}})\citenamefont{Schuetz, Brubaker, and
  Katzgraber}}]{schuetz:22}
\bibinfo{author}{\bibfnamefont{M.~J.~A.} \bibnamefont{Schuetz}},
  \bibinfo{author}{\bibfnamefont{J.~K.} \bibnamefont{Brubaker}},
  \bibnamefont{and} \bibinfo{author}{\bibfnamefont{H.~K.}
  \bibnamefont{Katzgraber}}, \emph{\bibinfo{title}{{Combinatorial Optimization
  with Physics-Inspired Graph Neural Networks}}}, \bibinfo{journal}{Nat. Mach.
  Intell.} \textbf{\bibinfo{volume}{4}}, \bibinfo{pages}{367}
  (\bibinfo{year}{2022}{\natexlab{a}}).

\bibitem[{\citenamefont{Duckworth and Zito}(2009)}]{duckworth:09}
\bibinfo{author}{\bibfnamefont{W.}~\bibnamefont{Duckworth}} \bibnamefont{and}
  \bibinfo{author}{\bibfnamefont{M.}~\bibnamefont{Zito}},
  \emph{\bibinfo{title}{{Large independent sets in random regular graphs}}},
  \bibinfo{journal}{Theoretical Computer Science}
  \textbf{\bibinfo{volume}{410}}, \bibinfo{pages}{5236} (\bibinfo{year}{2009}).

\bibitem[{\citenamefont{Schuetz
  et~al.}(2022{\natexlab{b}})\citenamefont{Schuetz, Brubaker, Zhu, and
  Katzgraber}}]{schuetz:22b}
\bibinfo{author}{\bibfnamefont{M.~J.~A.} \bibnamefont{Schuetz}},
  \bibinfo{author}{\bibfnamefont{J.~K.} \bibnamefont{Brubaker}},
  \bibinfo{author}{\bibfnamefont{Z.}~\bibnamefont{Zhu}}, \bibnamefont{and}
  \bibinfo{author}{\bibfnamefont{H.~K.} \bibnamefont{Katzgraber}},
  \emph{\bibinfo{title}{{Graph Coloring with Physics-Inspired Graph Neural
  Networks}}} (\bibinfo{year}{2022}{\natexlab{b}}),
  \bibinfo{note}{arXiv:2202.01606}.

\bibitem[{\citenamefont{Zheng et~al.}(2020)\citenamefont{Zheng, Ma, Wang, Zhou,
  Su, Song, Gan, Zhang, and Karypis}}]{zheng:20}
\bibinfo{author}{\bibfnamefont{D.}~\bibnamefont{Zheng}},
  \bibinfo{author}{\bibfnamefont{C.}~\bibnamefont{Ma}},
  \bibinfo{author}{\bibfnamefont{M.}~\bibnamefont{Wang}},
  \bibinfo{author}{\bibfnamefont{J.}~\bibnamefont{Zhou}},
  \bibinfo{author}{\bibfnamefont{Q.}~\bibnamefont{Su}},
  \bibinfo{author}{\bibfnamefont{X.}~\bibnamefont{Song}},
  \bibinfo{author}{\bibfnamefont{Q.}~\bibnamefont{Gan}},
  \bibinfo{author}{\bibfnamefont{Z.}~\bibnamefont{Zhang}}, \bibnamefont{and}
  \bibinfo{author}{\bibfnamefont{G.}~\bibnamefont{Karypis}},
  \emph{\bibinfo{title}{{{DistDGL: Distributed Graph Neural Network Training
  for Billion-Scale Graphs}}}} (\bibinfo{year}{2020}),
  \bibinfo{note}{(arXiv:2010.05337)}.

\bibitem[{\citenamefont{Leskovec}()}]{leskovec}
\bibinfo{author}{\bibfnamefont{J.}~\bibnamefont{Leskovec}},
  \emph{\bibinfo{title}{{{lectures on "Structure and models of real-world
  graphs and networks"}}}}, \bibinfo{note}{(Carnegie Mellon and Stanford
  University)}.

\bibitem[{\citenamefont{Monti et~al.}(2019)\citenamefont{Monti, Frasca, Eynard,
  Mannion, and Bronstein}}]{monti:19}
\bibinfo{author}{\bibfnamefont{F.}~\bibnamefont{Monti}},
  \bibinfo{author}{\bibfnamefont{F.}~\bibnamefont{Frasca}},
  \bibinfo{author}{\bibfnamefont{D.}~\bibnamefont{Eynard}},
  \bibinfo{author}{\bibfnamefont{D.}~\bibnamefont{Mannion}}, \bibnamefont{and}
  \bibinfo{author}{\bibfnamefont{M.}~\bibnamefont{Bronstein}},
  \emph{\bibinfo{title}{{{Fake news detection on social media using geometric
  deep learning}}}} (\bibinfo{year}{2019}), \bibinfo{note}{arXiv:1902.06673}.

\bibitem[{\citenamefont{Bronstein et~al.}(2021)\citenamefont{Bronstein, Bruna,
  Cohen, and Velickovic}}]{bronstein:21}
\bibinfo{author}{\bibfnamefont{M.~M.} \bibnamefont{Bronstein}},
  \bibinfo{author}{\bibfnamefont{J.}~\bibnamefont{Bruna}},
  \bibinfo{author}{\bibfnamefont{T.}~\bibnamefont{Cohen}}, \bibnamefont{and}
  \bibinfo{author}{\bibfnamefont{P.}~\bibnamefont{Velickovic}},
  \emph{\bibinfo{title}{{{Geometric Deep Learning: Grids, Groups, Graphs,
  Geodesics, and Gauges}}}} (\bibinfo{year}{2021}),
  \bibinfo{note}{(arXiv:2104.13478)}.

\bibitem[{\citenamefont{Fan et~al.}()\citenamefont{Fan, Shen, Nussinov, Liu,
  Sun, and Liu}}]{fan:21}
\bibinfo{author}{\bibfnamefont{C.}~\bibnamefont{Fan}},
  \bibinfo{author}{\bibfnamefont{M.}~\bibnamefont{Shen}},
  \bibinfo{author}{\bibfnamefont{Z.}~\bibnamefont{Nussinov}},
  \bibinfo{author}{\bibfnamefont{Z.}~\bibnamefont{Liu}},
  \bibinfo{author}{\bibfnamefont{Y.}~\bibnamefont{Sun}}, \bibnamefont{and}
  \bibinfo{author}{\bibfnamefont{Y.-Y.} \bibnamefont{Liu}},
  \emph{\bibinfo{title}{{{Finding spin glass ground states through deep
  reinforcement learning}}}}, \bibinfo{note}{arXiv:2109.14411}.

\bibitem[{\citenamefont{Gilmer et~al.}(2017)\citenamefont{Gilmer, Schoenholz,
  Riley, Vinyals, and Dahl}}]{gilmer:17}
\bibinfo{author}{\bibfnamefont{J.}~\bibnamefont{Gilmer}},
  \bibinfo{author}{\bibfnamefont{S.~S.} \bibnamefont{Schoenholz}},
  \bibinfo{author}{\bibfnamefont{P.~F.} \bibnamefont{Riley}},
  \bibinfo{author}{\bibfnamefont{O.}~\bibnamefont{Vinyals}}, \bibnamefont{and}
  \bibinfo{author}{\bibfnamefont{G.~E.} \bibnamefont{Dahl}}, in
  \emph{\bibinfo{booktitle}{Proceedings of the 34th International Conference on
  Machine Learning-Volume}} (\bibinfo{organization}{JMLR},
  \bibinfo{year}{2017}), vol.~\bibinfo{volume}{70}, p. \bibinfo{pages}{1263}.

\bibitem[{\citenamefont{Xu et~al.}(2019)\citenamefont{Xu, Weihua, Leskovec, and
  Jegelka}}]{xu:19}
\bibinfo{author}{\bibfnamefont{K.}~\bibnamefont{Xu}},
  \bibinfo{author}{\bibfnamefont{H.}~\bibnamefont{Weihua}},
  \bibinfo{author}{\bibfnamefont{J.}~\bibnamefont{Leskovec}}, \bibnamefont{and}
  \bibinfo{author}{\bibfnamefont{S.}~\bibnamefont{Jegelka}}, in
  \emph{\bibinfo{booktitle}{{International Conference on Learning
  Representations}}} (\bibinfo{year}{2019}).

\bibitem[{\citenamefont{Topping et~al.}(2021)\citenamefont{Topping,
  Di~Giovanni, Chamberlain, Dong, and Bronstein}}]{topping:21}
\bibinfo{author}{\bibfnamefont{J.}~\bibnamefont{Topping}},
  \bibinfo{author}{\bibfnamefont{F.}~\bibnamefont{Di~Giovanni}},
  \bibinfo{author}{\bibfnamefont{B.~P.} \bibnamefont{Chamberlain}},
  \bibinfo{author}{\bibfnamefont{X.}~\bibnamefont{Dong}}, \bibnamefont{and}
  \bibinfo{author}{\bibfnamefont{M.~M.} \bibnamefont{Bronstein}},
  \emph{\bibinfo{title}{{Understanding over-squashing and bottlenecks on graphs
  via curvature}}} (\bibinfo{year}{2021}), \bibinfo{note}{arXiv:2111.14522}.

\end{thebibliography}

\end{document}